# Visualizing and manipulating chiral interface states in a moiré quantum anomalous Hall insulator

Canxun Zhang[1,2,3,6], Tiancong Zhu[1,2,6]*, Salman Kahn[1,2,6], Tomohiro Soejima[1], Kenji Watanabe[4], Takashi Taniguchi[5], Alex Zettl[1,2,3], Feng Wang[1,2,3], Michael P. Zaletel[1,2], Michael F. Crommie[1,2,3]*

[1]Department of Physics, University of California, Berkeley, CA 94720, USA.
[2]Materials Sciences Division, Lawrence Berkeley National Laboratory, Berkeley, CA 94720, USA.
[3]Kavli Energy NanoScience Institute at the University of California, Berkeley and the Lawrence Berkeley National Laboratory, Berkeley, CA 94720, USA.
[4]Research Center for Electronic and Optical Materials, National Institute for Materials Science, 1-1 Namiki, Tsukuba 305-0044, Japan.
[5]Research Center for Materials Nanoarchitectonics, National Institute for Materials Science, 1-1 Namiki, Tsukuba 305-0044, Japan.
[6]These authors contributed equally: Canxun Zhang, Tiancong Zhu, Salman Kahn.
*Email: tiancongzhu@berkeley.edu; crommie@berkeley.edu.

**Abstract**

Moiré systems made from stacked two-dimensional materials host novel correlated and topological states that can be electrically controlled via applied gate voltages. We have used this technique to manipulate Chern domains in an interaction-driven quantum anomalous Hall insulator made from twisted monolayer–bilayer graphene (tMBLG). This has allowed the wavefunction of chiral interface states to be directly imaged using a scanning tunneling microscope (STM). To accomplish this tMBLG carrier concentration was tuned to stabilize neighboring domains of opposite Chern number, thus providing topological interfaces completely devoid of any structural boundaries. STM tip pulse-induced quantum dots were utilized to induce new Chern domains and thereby create new chiral interface states with *tunable* chirality at predetermined locations. Theoretical analysis confirms the chiral nature of observed interface states and enables the determination of the characteristic length scale of valley polarization reversal across neighboring tMBLG Chern domains. tMBLG is shown to be a useful platform for imaging the exotic topological properties of correlated moiré systems.

**Main**

Van der Waals stacks of two-dimensional (2D) atomic sheets provide a versatile platform for exploring topologically non-trivial phases of matter. The formation of moiré superlattices due to rotational misalignment and/or lattice mismatch induces energetically-narrow mini-bands[1-3] that inherit the large Berry curvature of the individual atomic layers.[4-6] Spontaneous valley polarization in the presence of strong electron–electron interactions can break the time-reversal (TR) symmetry, leading to a variety of topological states such as quantum anomalous Hall (QAH) insulators,[7-9] fractional Chern insulators,[10-12] and topological charge density waves.[13] The non-trivial topology and associated magnetic order of these moiré systems can be manipulated via electrical means without high magnetic field,[7,8,14,15] thus promising exciting new applications ranging from ultralow-power magnetic memories[8] to topological quantum computation.[16-18] The microscopic mechanisms at work in these materials and their ultimate performance in future applications, however, depend sensitively on spatial characteristics of chiral interface states that up to now have proved exceptionally difficult to visualize due to extrinsic factors such as structural defects.

We have utilized scanning tunneling microscopy and spectroscopy (STM/STS) to manipulate Chern domains and to directly visualize the wavefunction of chiral interface states in a moiré QAH insulator made from twisted monolayer–bilayer graphene (tMBLG). Neighboring tMBLG domains were induced to exhibit opposite Chern number by tuning the electron density in our devices via electrostatic gating. Spatially-resolved STS was used to visualize topological phase transitions across the interfaces between such domains and to directly probe the local electronic structure of resulting one-dimensional (1D) chiral modes. We have discovered that these 1D chiral states can be spatially displaced on demand via global back-gating. Moreover, we are able to *create* new interface states with predetermined chirality and position via tip pulse-induced quantum dot (QD) formation.[19] Comparison of our data to an effective theoretical model confirms that chiral interface states arise in tMBLG due to a reversal of valley polarization across Chern domains and allows us to determine the characteristic width of a Chern domain wall, a length scale that has significant implications for potential device applications.

**Density control of local Chern number for tMBLG QAH states**

Figure 1a shows a sketch of our experiment which incorporates a tMBLG device within an STM measurement geometry. Here a Bernal-stacked bilayer graphene is placed on top of a monolayer graphene with a twist angle $\theta \approx 1.25°$ between them, and the stack is supported by a hexagonal boron nitride (hBN) substrate placed on an Si/SiO$_2$ wafer that functions as the back-gate (Methods). The monolayer–bilayer rotational misalignment generates a moiré superlattice of wavelength $l_M = 11.3$ nm (Extended Data Fig. 1a) that folds dispersive electronic bands into energetically-narrow mini-bands, each accommodating four electrons per moiré unit cell due to spin and valley degeneracies (Fig. 1b).[20,21] Tuning the back-gate voltage $V_G$ allows us to dope the graphene stack with charge carriers that partially occupy the moiré mini-band electronic states (Extended Data Fig. 1b). At integer filling factors $v$ (defined as the average number of electrons per moiré unit cell referenced to charge neutrality; see Supplementary Note 1) strong electron–electron interactions can induce correlated insulating states that lift the spin/valley degeneracy. Figure 1c shows a representative d$I$/d$V$ spectrum obtained at $V_G = 41.6$ V ($v = 3$) where a charge gap with nearly vanishing d$I$/d$V$ appears around $V_{Bias} = 0$ mV (i.e., the Fermi energy $E_F$), flanked by a "lower band" (LB) peak and an "upper band" (UB) peak.

This $v = 3$ insulating state exhibits non-trivial topological behavior as shown by the QAH effect reported in previous transport measurements.[8] We are able to directly observe its topological properties via STM by performing gate-dependent d$I$/d$V$ spectroscopy in an applied out-of-plane magnetic field $B$.[22] Figure 1d shows a d$I$/d$V$ density plot for $B = 0.0$ T where the $v = 3$ gap is marked by a white arrow. Under application of a small magnetic field ($B = 0.4$ T), this gap evolves into *two separate gaps* that exhibit similar spectroscopic features but are situated at different electron densities, one above and one below $v = 3$ (orange and green arrows in Fig. 1e). These two gaps split further away from $v = 3$ with increasing $B$, as shown by the V-shaped dark region in the d$I$/d$V$ density plot obtained at $V_{\text{Bias}} = 0$ mV (Fig. 1f; see Extended Data Fig. 2 for additional details). Such linear scaling between gap position and $B$-field is indicative of a QAH insulating state whose Chern number is either $C = +2$ (for the $v > 3$ branch) or $C = -2$ (for the $v < 3$ branch) as extracted from the Středa formula $\frac{\Delta n}{\Delta B} = C \frac{e}{h}$.[23] This separation of $C = \pm 2$ states allows us to independently access different Chern states in a small $B$-field by tuning the electron density via electrostatic gating, thus providing electrical control over local Chern number in our tMBLG device.

**Topological phase transition and 1D chiral states at Chern domain interfaces**

Electrical control over the tMBLG Chern number provides us with a unique opportunity to stabilize neighboring insulating domains with opposite local Chern values and to explore the topological phase transition across them in regions exhibiting charge inhomogeneity. Figure 2a shows an STM topographic image of such an area in tMBLG as seen by the d$I$/d$V$ density plots of Fig. 2b for $B = 0.0$ T obtained at locations *1–3*. Spatial inhomogeneity in the charge density is reflected in the $v = 3$ gap (white arrows) which occurs at different $V_G$ values for the three different locations. Application of an out-of-plane field of $B = 0.4$ T causes the $v = 3$ gap to evolve into two gaps of opposite Chern number as shown in Fig. 2c, with the charge inhomogeneity causing the gaps to occur at different gate voltages at different locations. At $V_G = $ 42.0 V (dashed line in Fig. 2c) location *1* acquires the lower gap (left panel), indicating that it is in a $C = -2$ insulating domain, whereas location *3* acquires the higher gap (right panel), indicating that it is in a $C = +2$ insulating domain.

To visualize tMBLG local electronic structure between insulating domains of opposite Chern number, Fig. 2f shows d*I*/d*V* spectra obtained along the white dashed line in Fig. 2a (representative point spectra are shown in Fig. 2d). A gap feature is clearly seen in the far left ($x < 80$ nm) and far right ($x > 160$ nm) regions, along with the LB and UB peaks. The spectra in the middle region (80 nm $< x <$ 160 nm), however, exhibit very different behavior. The LB peak persists (with reduced intensity) but the UB peak can no longer be clearly resolved. Moreover, an increase in d*I*/d*V* signal emerges at $V_{Bias} = 0$ mV (i.e., $E_F$) that closes the insulating gap (this can be seen in the middle d*I*/d*V* spectrum of Fig. 2d as well as Extended Data Fig. 3c). These data are the signature of a real-space topological phase transition that starts in a $C = -2$ insulating domain and progresses through a conducting interface to a $C = +2$ insulating domain. It is important to note that the tMBLG region at the interface between these two Chern domains is unblemished and exhibits no structural defects.

The presence of a conducting region at the Chern domain interface is consistent with the emergence of 1D chiral modes between QAH insulators having different Chern numbers. The microscopic wavefunction of such modes can be seen in Fig. 2e through a d*I*/d*V* map taken at $V_{Bias} = 0$ mV of the same area as in Fig. 2a (for $V_G = 42.0$ V and $B = 0.4$ T; see Methods). The $C = -2$ and $C = +2$ domains in Fig. 2e appear as dark regions with nearly vanishing d*I*/d*V* whereas the bright stripe in the middle is a visualization of the 1D chiral interface modes (which extend diagonally from the lower left to the upper right). Similar spatially-defined topological phase transitions (accompanied by chiral interface states) were observed in numerous different tMBLG areas having inhomogeneous electron density (see Extended Data Fig. 4). 2D Gaussian fit profiles made for zero-bias d$I$/d$V$ maps at four different topological interfaces result in a full width at half maximum (FWHM) for tMBLG chiral interface modes equal to 57 ± 10 nm (Methods; see Extended Data Fig. 5 and Extended Data Table 1).

**Manipulating chiral interface states via electrostatic gating**

The emergence of Chern domains and chiral interface states in tMBLG that are sensitive to local electron density (but independent of structural boundaries) allows flexible manipulation of their spatial location via electrostatic back-gating. Figure 3 shows a series of zero-bias d$I$/d$V$ maps of the same chiral interface modes shown in Fig. 2e as $V_G$ is gradually decreased from 42.5 V to 41.5 V (the region on the right has a higher local electron density $n$ than the left region). At $V_G$ = 42.5 V (Fig. 3a) the electron density is above $v$ = 3 for the entire scanned area. The darkest region on the left side is in the $C$ = +2 insulating state (i.e., d$I$/d$V$ ≈ 0) whereas the region on the right side has some additional electron doping that results in d$I$/d$V$ > 0. Decreasing $V_G$ (Fig. 3b) reduces the local electron density everywhere, causing the $C$ = +2 insulating domain to shift toward the right (i.e., according to the gradient shown in Fig. 2c). At $V_G$ = 42.1 V the left region switches to a $C$ = –2 insulating state and the right region exhibits a $C$ = +2 state, and bright 1D states remain at the interface between them (Fig. 3c). Further decreasing $V_G$ leads to expansion of the $C$ = –2 domain and reduction of the $C$ = +2 domain, moving the interface states continuously from left to right (Fig. 3d, e). Finally at $V_G$ = 41.5 V the entire scanned area lies below $v$ = 3 with the darkest region on the right side representing a $C$ = –2 insulating state and the region to its left being slightly hole-doped (Fig. 3f). This interface state manipulation process is completely reversible as $V_G$ is increased back to 42.5 V (see Extended Data Fig. 6).

**Creating new interface states with reversible chirality**

We are able to *create* new chiral interface states by design by inducing charge inhomogeneity through STM tip pulse-induced QDs. Figure 4a shows an area of a tMBLG device having relatively low variation of local electron density $n$ in its pristine state (see Extended Data Fig. 7b for additional details). To alter the density profile we applied a bias voltage pulse of 5 V at the location marked by the red dot in Fig. 4a while holding the gate voltage fixed at $V_G = -2$ V (Fig. 4b). This process changes the charge state of defects in the hBN substrate, resulting in an n-type (i.e., electron-doped) graphene QD centered at the site of the bias pulse[19] (see Supplementary Note 2 and Extended Data Fig. 7c for additional details). Under these conditions the tMBLG electron density decreases away from the QD center, thus causing a density gradient that induces a Chern domain interface and chiral interface states near the QD edge as described previously in Fig. 2.

Chiral interface states created in this way can be identified using STS as shown in Fig. 4c. Under application of $B = 0.4$ T and $V_G = 43.0$ V the location marked "*1*" in Fig. 4a is found to be in the gapped $C = +2$ state (Fig. 4c left panel) while the location marked "*3*" (which has lower local $n$) is found to be in the gapped $C = -2$ state (Fig. 4c right panel). The location between these two points (marked "*2*" in Fig. 4a) shows a collapse of the energy gap and thus indicates the position of *new* chiral interface states (Fig. 4c middle panel) (see Extended Data Fig. 8a–c for additional details). Such chiral states propagate unidirectionally according to the Chern number difference across the interface, and so one can deduce that electrons must flow downward along this interface based on the Chern numbers of the neighboring domains (Fig. 4d).

The chirality (and thus the direction of electron flow) for these tailor-made interface states can be *flipped* by reversing the polarity of the QD. To accomplish this we applied a bias voltage pulse of 5 V to the same location in Fig. 4a as before while holding $V_G = +2$ V as shown in Fig. 4e. This erased the previous QD and induced a p-type (i.e., hole-doped) graphene QD in the same area, thus *reversing* the electron density gradient in regions *1–3*. Under application of $B$ = 0.4 T and $V_G$ = 45.0 V, a gapped $C = -2$ domain is now observed at location *1* (Fig. 4f left panel) and a gapped $C = +2$ domain is observed at location *3* (Fig. 4f right panel). Location *2* (Fig. 4f middle panel) shows a collapse of the energy gap and thus marks the location of newly formed chiral interface modes, but now with the difference that the domain Chern numbers are reversed and thus the electron flow is flipped (Fig. 4g). The chirality of artificially-fabricated interface modes can thus be *switched* (see Extended Data Fig. 8d–f for additional details).

**Theoretical model and discussion**

The behavior that we observe for Chern domains and chiral interface states in tMBLG can be understood by analyzing the origin of non-trivial topology of interaction-driven insulating states. The tMBLG moiré mini-band investigated in our experiment (marked red in Fig. 1b) is four-fold degenerate, with each sub-band originating from an unfolded $K_+$ or $K_-$ valley having a non-zero Chern number of $c(K_\pm) = \pm 2$ due to the large Berry curvature inherited from the constituent graphene layers.[20,21] At $v = 3$ strong Coulomb exchange interactions induce spontaneous polarization in the spin–valley space, with three of the four spin- and valley-resolved sub-bands becoming filled (corresponding to the experimental LB peak shown in Fig. 1c) and the remaining one being empty (corresponding to the experimental UB peak shown in Fig. 1c). This leads to a TR symmetry-broken insulating state with a total Chern number $C \neq 0$ due to unbalanced electron occupation between the two valleys (Fig. 1g).[8,22] Double occupancy in the $K_+$ valley and single occupancy in the $K_-$ valley results in a gapped $C = +2$ state (Fig. 1g top panel) whereas the opposite configuration leads to a $C = -2$ state (Fig. 1g bottom panel).

Neighboring $C = +2$ and $C = -2$ insulating domains as observed in our experiment correspond to opposite valley occupancies, and so valley polarization reversal is expected across the domain interface. The simplest scenario for such a transition is illustrated in Fig. 2h. As one moves from the $C = -2$ domain (location *1*) to the $C = +2$ domain (location *3*), the empty $K_+$ sub-band smoothly shifts downward in energy and becomes occupied, while a filled $K_-$ sub-band smoothly shifts upward and becomes depleted. This qualitative picture leads to a closing (location *2*) and reopening (location *3*) of the correlation gap across the domain interface, as observed experimentally (Fig. 2f). At the interface both a $K_+$ sub-band and a $K_-$ sub-band are partially occupied which leads to 1D conducting states at $E_F$.

Such a picture of valley polarization reversal allows us to construct a quantitative model of tMBLG Chern domain interfaces that enables the extraction of useful physical parameters from the chiral interface states observed in our experiment. Here the moiré mini-band is described by a tight-binding model derived from a continuum Hamiltonian,[20,21] and the correlation-induced valley polarization and reversal is modelled using a spatially-varying energy offset (see Supplementary Note 3). In the bulk of each insulating domain we assume an energy separation of $2\mathcal{E}_0 = 30$ meV between occupied and unoccupied electronic states as extracted from the experimental d$I$/d$V$ (Fig. 2d, f). Starting from a $C = -2$ domain and moving toward a $C = +2$ domain, the downward-shifting $K_+$ sub-band is first centered at $+\mathcal{E}_0$ (location *1* in Fig. 2h) and ends up centered at $-\mathcal{E}_0$ (location *3*). As the interface is crossed the energy offset shifts linearly from $+\mathcal{E}_0$ to $-\mathcal{E}_0$ over a valley domain wall width $= \xi$ that depends on details of the electron–electron interaction and the local carrier density profile. The energy offset for the upward-shifting $K_-$ sub-band has the opposite spatial profile. Solving for the electronic eigenstates of this model yields four 1D branches that span the bulk gap and disperse unidirectionally (Extended Data Fig. 9a, c), thus confirming the emergence of chiral interface modes residing between the $C = -2$ and $C = +2$ insulating domains.

Using this model we are able to reproduce the spatially-dependent topological phase transition that we observe experimentally as the interface is crossed and to explain the origin of the different features seen in our STM spectroscopy. Figure 2g shows the resulting theoretical local density of states (LDOS) across the domain interface for $v = 3$ and $\xi = 125$ nm. Here $\xi$ is the only parameter not fixed by experiment and $\xi = 125 \pm 20$ nm yields the best fit to our experimental data (Fig. 2f) (see Supplementary Note 3 for additional details). The calculated LDOS reproduces many of the qualitative features observed in the experiment. The theoretical LB feature at $E = -\mathcal{E}_0$ in Fig. 2g, for example, persists over the entire region from one domain to the next but has reduced intensity near the interface (i.e., for 80 nm $< x <$ 160 nm). The reduction in LB spectral intensity is seen to arise due to one of the three filled sub-bands shifting to higher energy as a result of the valley polarization reversal at the Chern domain interface, consistent with the simple qualitative picture in Fig. 2h. The theoretical UB peak at $E = +\mathcal{E}_0$ is completely suppressed at the interface and shifts to $E = 0$ (equivalent to $E_F$). The downward-shifting unoccupied sub-band and the upward-shifting occupied sub-band are seen to merge at $E = 0$ to create 1D chiral modes. The spatial dependence of the theoretical LDOS map at $E = 0$ (Extended Data Fig. 3d) qualitatively reproduces the experimental d$I$/d$V$ map at $E = E_F$ (Fig. 2e, Extended Data Fig. 3a) and yields a FWHM of the 1D chiral modes of $w = 45$ nm, in reasonable agreement with the experimental value of $w = 57 \pm 10$ nm.

Our combined experimental and theoretical results suggest that chiral interface states arise in tMBLG due to a gradual transition of the valley polarization across different Chern domains, and we are able to extract a domain wall width of $\xi = 125 \pm 20$ nm that characterizes the length scale of valley reversal in our tMBLG devices. This width, in conjunction with the width of the chiral interface state wavefunction, puts a lower bound on the spacing between chiral channels in future hypothetical devices utilizing tMBLG in order to prevent inter-channel electronic hybridization. Our treatment suggests the possibility of controlling the width and dispersion of 1D chiral states in tMBLG by engineering the valley domain wall width $\xi$ (see example in Extended Data 9). Further reduction of the chiral state width could potentially be realized by creating devices with patterned local-gate structures designed to achieve sharper $n$-gradients.[24]

In conclusion, our ability to manipulate tMBLG Chern domains and to visualize the resulting 1D chiral interface states provides a new method to locally control and characterize topologically non-trivial moiré systems via scanned probe techniques. Direct access to the microscopic wavefunction of tMBLG chiral interface states allows us to characterize the fundamental length scale of these novel electronic states, thus providing a means to determine the ultimate limits on device miniaturization involving topologically-derived chiral channels and to test theoretical models of strong correlation effects in interaction-driven Chern insulators. Our new capability of creating 1D conducting channels with desired spatial location and chirality can potentially be utilized to build electrically-tunable chiral networks that hold promise for dissipation-free performance[25] and that should facilitate exploration of new exotic topological phenomena.[16,17,26]

**Methods**

**Sample preparation.** tMBLG samples were prepared using the "flip-chip" method[27] followed by a forming-gas anneal.[28,29] Electrical contacts were made by evaporating Cr/Au (5

nm/60 nm) through a silicon nitride shadow-mask onto the heterostructure. The sample surface cleanliness was confirmed using contact-AFM prior to STM measurements. Samples were annealed at 300 °C overnight in ultrahigh vacuum before insertion into the low-temperature STM stage.

**STM/STS measurements.** All STM/STS measurements were performed in a commercial CreaTec LT-STM held at $T = 4.7$ K using tungsten (W) tips. STM tips were prepared on a Cu(111) surface and calibrated against the Cu(111) Shockley surface state before STS measurements on tMBLG to avoid tip artifacts. d$I$/d$V$ spectra were recorded using standard lock-in techniques with a small bias modulation of $\Delta V_{RMS} = 1$mV at 613 Hz. d$I$/d$V$ maps were collected by first stabilizing the tip at a higher bias at each position, then bringing it closer to the sample surface by a fixed offset $\Delta Z$ for measurement (thus allowing the influence of structural corrugation to be minimized). All STM images were edited using WSxM software.[30]

**Gaussian fitting of chiral interface state images.** For fitting purposes we assumed that the chiral interface state LDOS is uniform *along* the domain interface and follows a Gaussian line shape normal to the interface in the plane. We also include a phenomenological constant background term to account for residual d$I$/d$V$ in the $C = \pm 2$ insulating domains (likely due to slight doping and/or instrumental broadening). The 1D chiral interface state LDOS can thus be expressed as

$$D(x,y) = D_0 + A \exp\left(-4 \ln 2 \left(\frac{(x - x_0)\cos\theta + (y - y_0)\sin\theta}{w}\right)^2\right)$$

where $D_0$ is the constant background, $A$ is the maximum intensity of the Gaussian peak, $(x_0, y_0)$ defines the interface, $\theta$ is the angle between the normal to the interface and the $x$ axis, and $w$ is the Gaussian FWHM. Extended Data Fig. 5a, c, e, g shows experimental d$I$/d$V$ maps obtained in four different tMBLG areas (#1–4) and Extended Data Fig. 5b, d, f, h shows the corresponding fits. The extracted FWHM for different datasets are summarized in Extended Data Table 1 and have an average width of $w = 57 \pm 10$ nm $\approx 5l_M$ where $l_M$ is the moiré wavelength. $w$ exhibits some slight variation for different electron density gradients and $B$-fields, but the exact relationship is not yet determined beyond the parameter regime presented here.

**Data Availability**

The data that support the plots within this paper and the findings of this study are available from the corresponding authors upon request.

**Code Availability**

The computer codes that support the plots within this paper and the findings of this study are available from the corresponding authors upon request.

**Acknowledgements**

The authors thank Yuan-Ming Lu for helpful discussion and Samuel Stolz for technical support. This research was supported by the Director, Office of Science, Office of Basic Energy Sciences, Materials Sciences and Engineering Division of the US Department of Energy, under contract number DE-AC02-05CH11231, within the van der Waals Heterostructures program KCWF16 (STM measurement and instrumentation). For graphene characterization we used the Molecular Foundry at LBNL, which is funded by the Director, Office of Science, Office of Basic Energy Sciences, Scientific User Facilities Division, of the US Department of Energy under Contract No. DE-AC02-05CH11231. Support was also provided by National Science Foundation Award DMR-1807233 (device fabrication, image analysis). K.W. and T.T. acknowledge support from JSPS KAKENHI (Grant Numbers 20H00354, 21H05233 and 23H02052) and World Premier International Research Center Initiative (WPI), MEXT, Japan (hBN crystal synthesis and characterization). C.Z. acknowledges support from a Kavli ENSI Philomathia Graduate Student Fellowship. T.S. acknowledges fellowship support from the Masason Foundation.

**Author Contributions**

C.Z., T.Z., and M.F.C. initiated and conceived the research. C.Z. and T.Z. carried out STM/STS measurements and analyses. M.F.C. supervised STM/STS measurements and analyses. S.K. prepared gate-tunable devices. A.Z., F.W. and M.F.C. supervised device preparations. T.S. performed theoretical analysis and calculations. M.P.Z. supervised theoretical analysis and calculations. T.T. and K.W. provided the hBN crystals. C.Z., T.Z., and M.F.C. wrote the manuscript with the help of all authors. All authors contributed to the scientific discussion.

**Competing Interests**

The authors declare no competing interests.

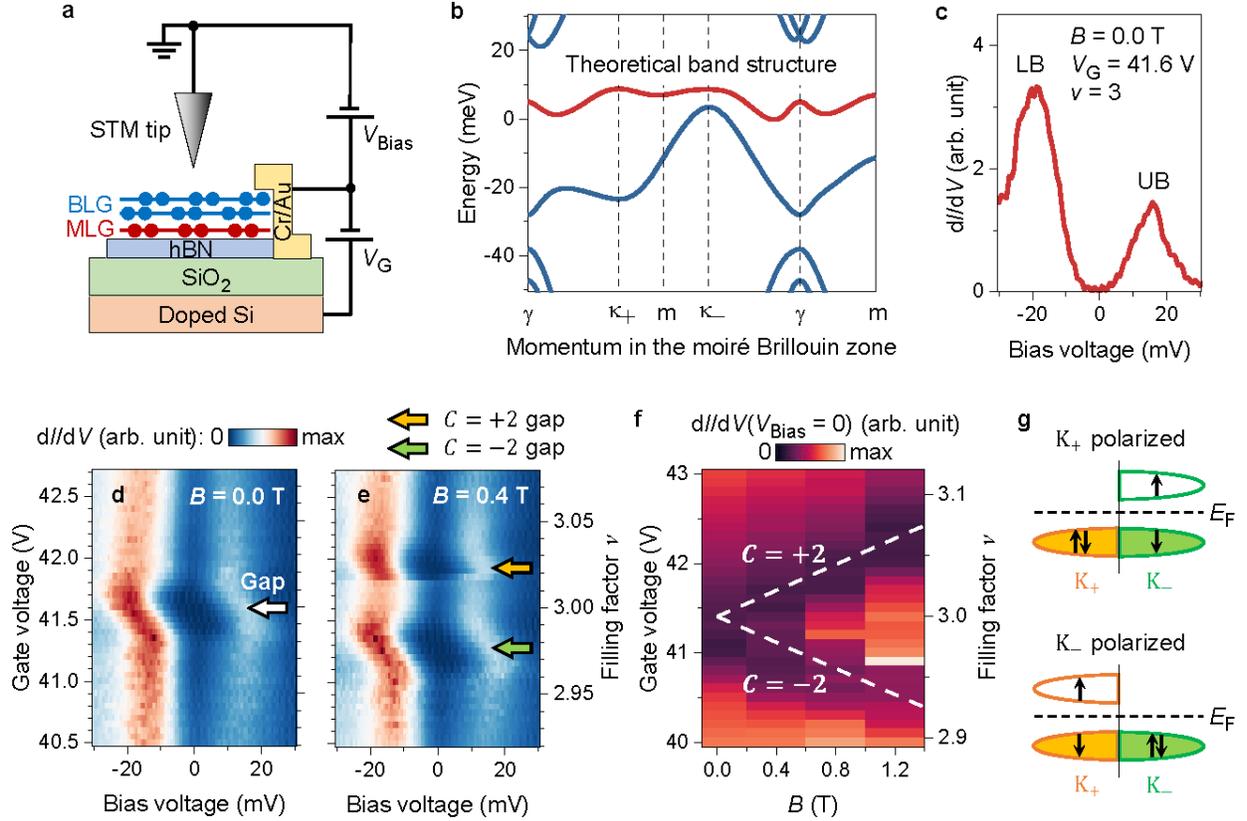

**Figure 1: Local characterization of gate-tunable Chern number for tMBLG QAH states. a**, Sketch of tMBLG device in STM/STS measurement geometry. MLG = monolayer graphene, BLG = Bernal-stacked bilayer graphene (there is a twist angle between MLG and BLG). **b**, Single-particle band structure of $\theta = 1.25°$ tMBLG along the high symmetry directions of the moiré Brillouin zone. For clarity only mini-bands from the unfolded $K_+$ valley are shown. **c**, $dI/dV$ spectrum obtained in an area with a local twist angle of 1.25° for $B = 0.0$ T and $V_G = 41.6$ V ($\nu = 3$). LB = lower band, UB = upper band. This is the same area as shown in Extended Data Fig. 1. **d**, Gate-dependent $dI/dV$ density plot for tMBLG near $\nu = 3$ ($B = 0.0$ T). The white arrow indicates the energy gap. **e**, Same as **d**, but at $B = 0.4$ T. Orange and green arrows indicate $C = +2$ and $C = -2$ gaps. **f**, $dI/dV(V_{Bias} = 0)$ as a function of $V_G$ and $B$-field for tMBLG (see Extended Data Fig. 2a–d for additional details). Dashed lines show the Středa formula for $C = \pm 2$. **g**, Schematic of tMBLG band filling for the unfolded $K_+$ valley-polarized state (top) and $K_-$ valley-polarized state (bottom) at $\nu = 3$. Arrows represent electron spin. Experimental spectroscopy parameters: modulation voltage $\Delta V_{RMS} = 1$ mV; setpoint $V_{Bias} = -60$ mV, $I_0 = 0.5$ nA. All measurements taken at $T = 4.7$ K.

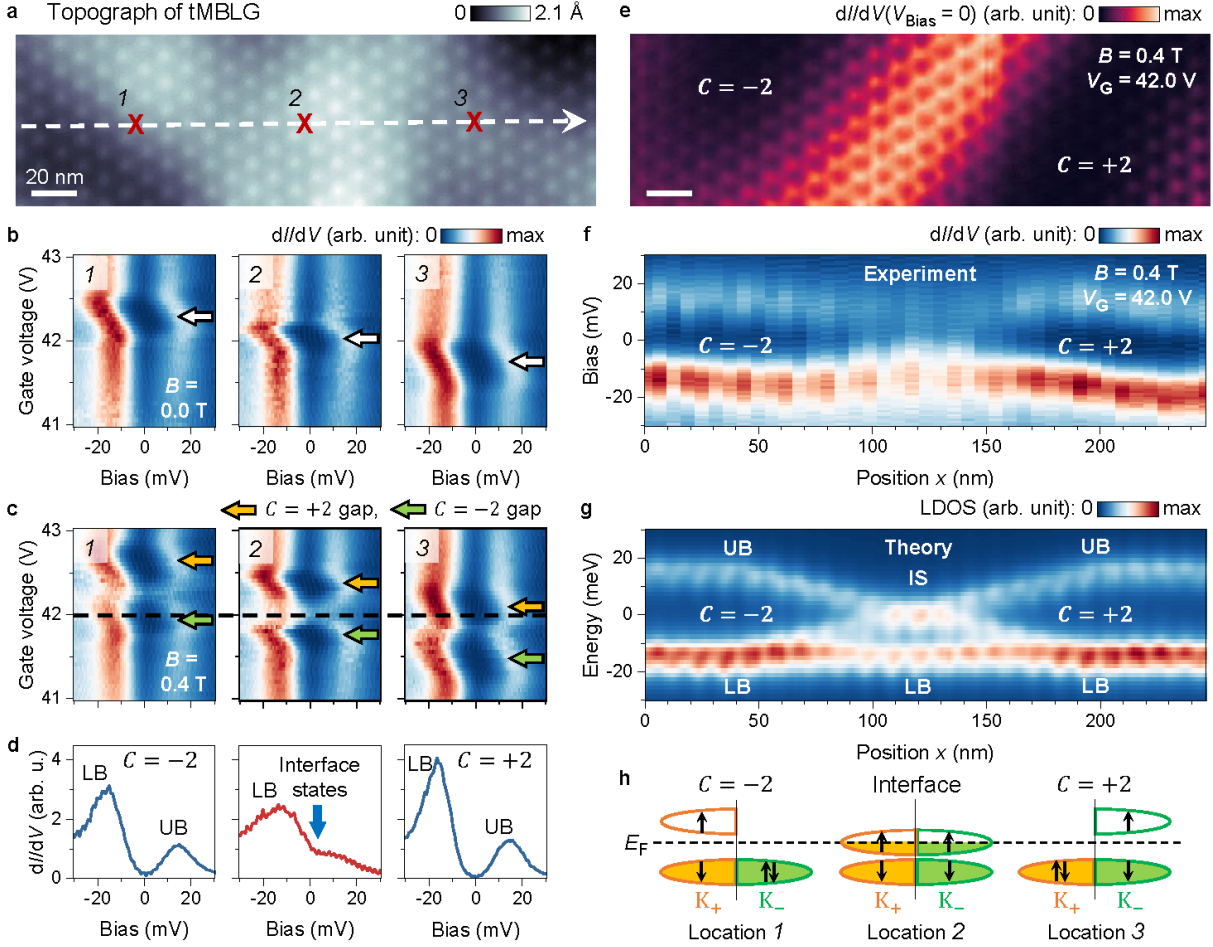

**Figure 2: Topological phase transition and 1D chiral states at a Chern domain interface. a**, STM topographic image of a tMBLG area having inhomogeneous electron density ($V_{Bias} = -300$ mV, setpoint $I_0 = 0.2$ nA). **b**, Gate-dependent $dI/dV$ density plots for $B = 0.0$ T obtained at locations *1*, *2*, and *3* marked in **a**. White arrows indicate the $\nu = 3$ energy gap. **c**, Same as **b**, but for $B = 0.4$ T. Orange and green arrows indicate $C = +2$ and $C = -2$ gaps. **d**, $dI/dV$ spectra for $B = 0.4$ T and $V_G = 42.0$ V taken at locations *1*, *2*, and *3* of **a**. LB = lower band, UB = upper band. **e**, $dI/dV$ map of the same area as **a** for $B = 0.4$ T, $V_G = 42.0$ V and $V_{Bias} = 0$ mV. Bright region shows chiral interface state density. **f**, $dI/dV$ density plot for $B = 0.4$ T and $V_G = 42.0$ V obtained along the white dashed line in **a**. **g**, Density plot of theoretical LDOS calculated for a Chern domain wall width of $\xi = 125$ nm (LB = lower band, UB = upper band, IS = interface states) (compare to **f**). **h**, Schematic of band diagram for topological phase transition across a tMBLG Chern domain interface. Arrows represent electron spin. Experimental spectroscopy parameters: modulation voltage $\Delta V_{RMS} = 1$ mV; setpoint $V_{Bias} = -60$ mV, $I_0 = 0.5$ nA for **b**, **c**; setpoint $V_{Bias} = -300$ mV, $I_0 = 0.2$ nA and tip height offset $\Delta Z = -0.075$ nm for **d**–**f**.

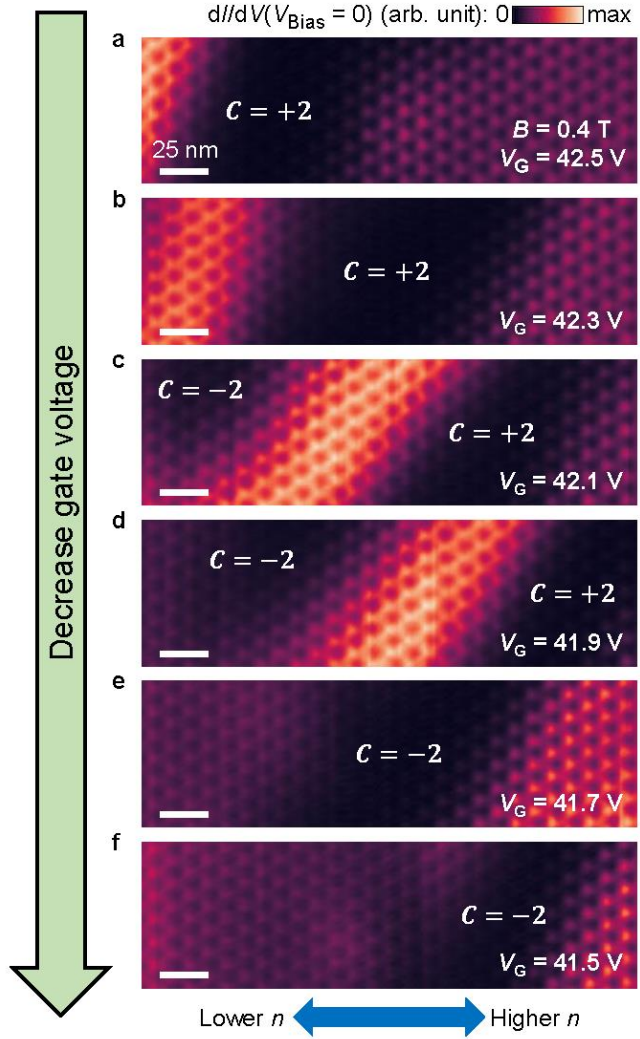

**Figure 3: Manipulating Chern domains and chiral interface states via back-gating. a–f**, d$I$/d$V$ maps of the same area shown in Fig. 2a at $B = 0.4$ T and $V_{Bias} = 0$ mV for **a** $V_G = 42.5$ V, **b** $V_G = 42.3$ V, **c** $V_G = 42.1$ V, **d** $V_G = 41.9$ V, **e** $V_G = 41.7$ V, and **f** $V_G = 41.5$ V. Modulation voltage $\Delta V_{RMS} = 1$ mV; setpoint $V_{Bias} = -300$ mV, $I_0 = 0.2$ nA; tip height offset $\Delta Z = -0.075$ nm.

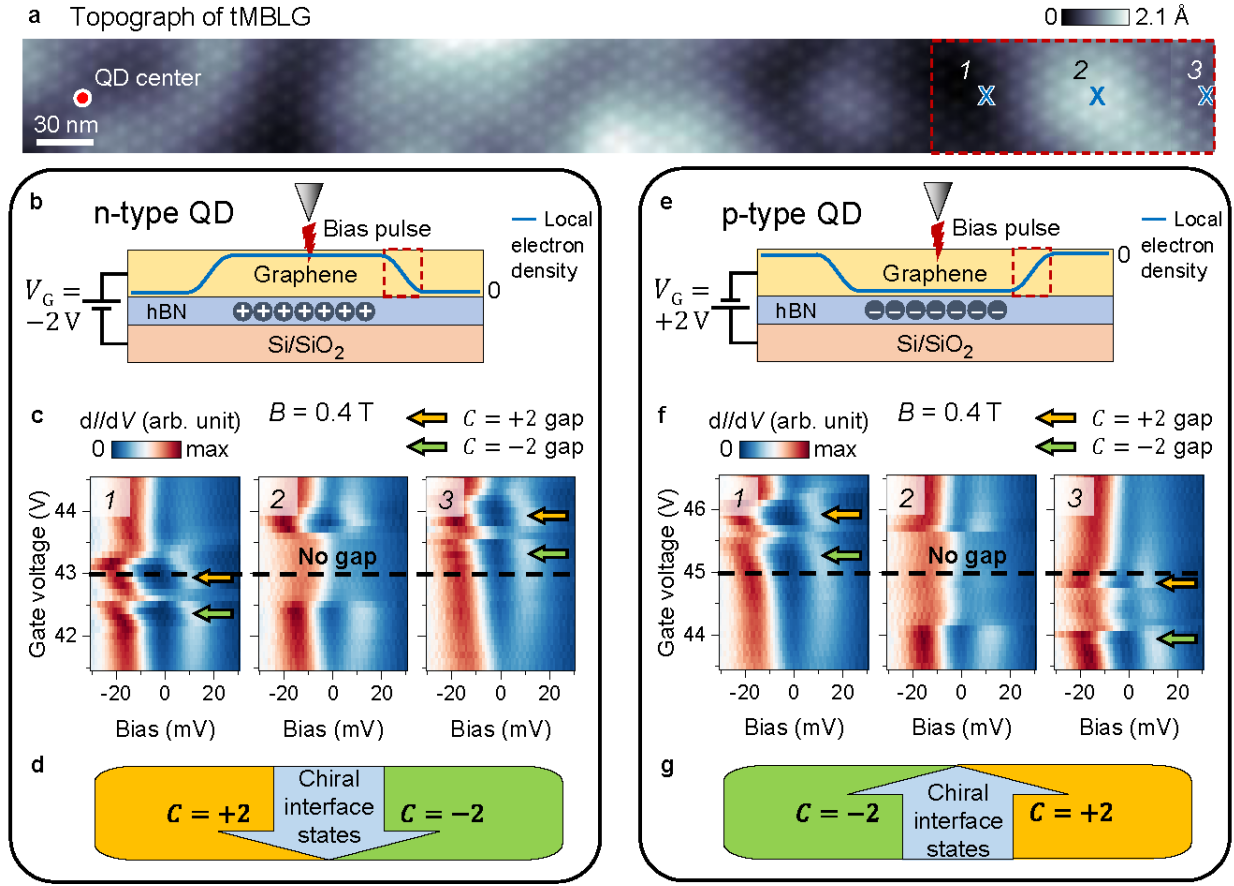

**Figure 4: Creating chiral interface states with tunable chirality via STM tip pulse-induced QDs. a**, STM topographic image of a tMBLG area after using an STM tip pulse to create a quantum dot (QD) ($V_{Bias} = –1$ V, setpoint $I_0 = 0.01$ nA). **b**, Schematic of the process of inducing an n-type QD. The blue curve signifies the resulting graphene QD local electron density. The red dashed box highlights the region with the largest density gradient and represents the boxed region in **a**. **c**, d$I$/d$V$ density plots for $B = 0.4$ T obtained at locations *1*, *2*, and *3* marked in **a** after creation of an n-type QD centered at the red dot in **a** (modulation voltage $\Delta V_{RMS} = 1$ mV; setpoint $V_{Bias} = –60$ mV, $I_0 = 0.5$ nA). Orange and green arrows indicate $C = +2$ and $C = –2$ energy gaps. **d**, Illustration representing the region encircled by the red dashed box in **a** for the case of an n-type QD with $B = 0.4$ T and $V_G = 43.0$ V (i.e., corresponding to the horizontal dashed line in **c**). Blue arrow shows direction of electron flow for chiral interface states. **e**, Schematic of the process of inducing a p-type QD. **f**, Same as **c**, but after creation of a p-type QD centered at the red dot in **a**. **g**, Same as **d** except for the case of a p-type QD with $B = 0.4$ T and $V_G = 45.0$ V (i.e., corresponding to the horizontal dashed line in **f**). Note that direction of blue arrow is reversed compared to **d**.

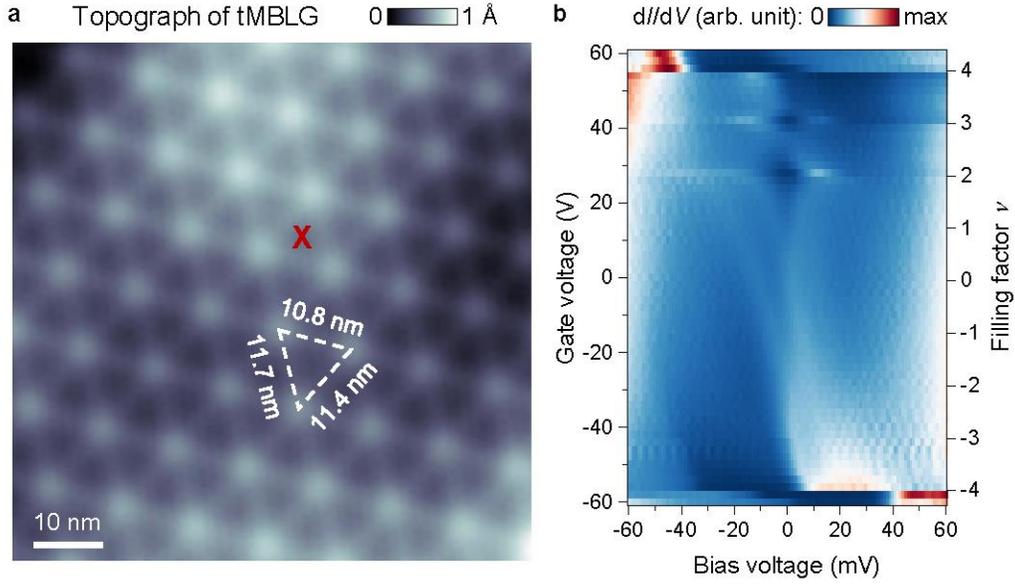

**Extended Data Figure 1: Basic STM/STS characterization of gate-tunable tMBLG. a**, STM topographic image of a tMBLG device surface area ($V_{Bias} = -200$ mV, setpoint $I_0 = 2$ nA). The average moiré wavelength of $l_M = 11.3$ nm corresponds to a local twist angle of $\theta = 1.25°$ (see Supplementary Note 1). **b**, Gate-dependent $dI/dV$ density plot over the full range $-70$ V $\leq V_G \leq 70$ V ($-4.3 \leq \nu \leq 4.3$) obtained for STM tip at the location marked in **a** (modulation voltage $\Delta V_{RMS} = 1$ mV; setpoint $V_{Bias} = -100$ mV, $I_0 = 2$ nA). Charge gaps emerging at $\nu = 2, 3$ signify the formation of correlated insulating states. All measurements taken at $T = 4.7$ K.

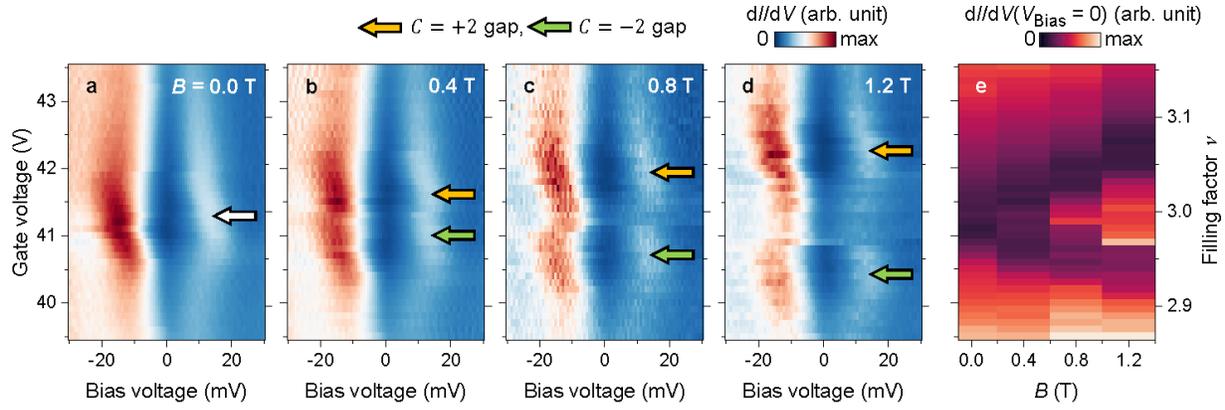

**Extended Data Figure 2: Topological behavior of tMBLG QAH states in an out-of-plane $B$-field. a–d**, Gate-dependent d$I$/d$V$ density plots near $v = 3$ for **a** $B = 0.0$ T, **b** $B = 0.4$ T, **c** $B = 0.8$ T, and **d** $B = 1.2$ T (modulation voltage $\Delta V_{RMS} = 1$ mV; setpoint $V_{Bias} = -75$ mV, $I_0 = 0.2$ nA). The white arrow indicates the energy gap for $B = 0.0$ T, and orange and green arrows indicate $C = +2$ and $C = -2$ gaps for finite $B$. **e**, d$I$/d$V$($V_{Bias} = 0$) as a function of $V_G$ and $B$-field extracted from **a–d** (same data as in Fig. 1f but without dashed lines).

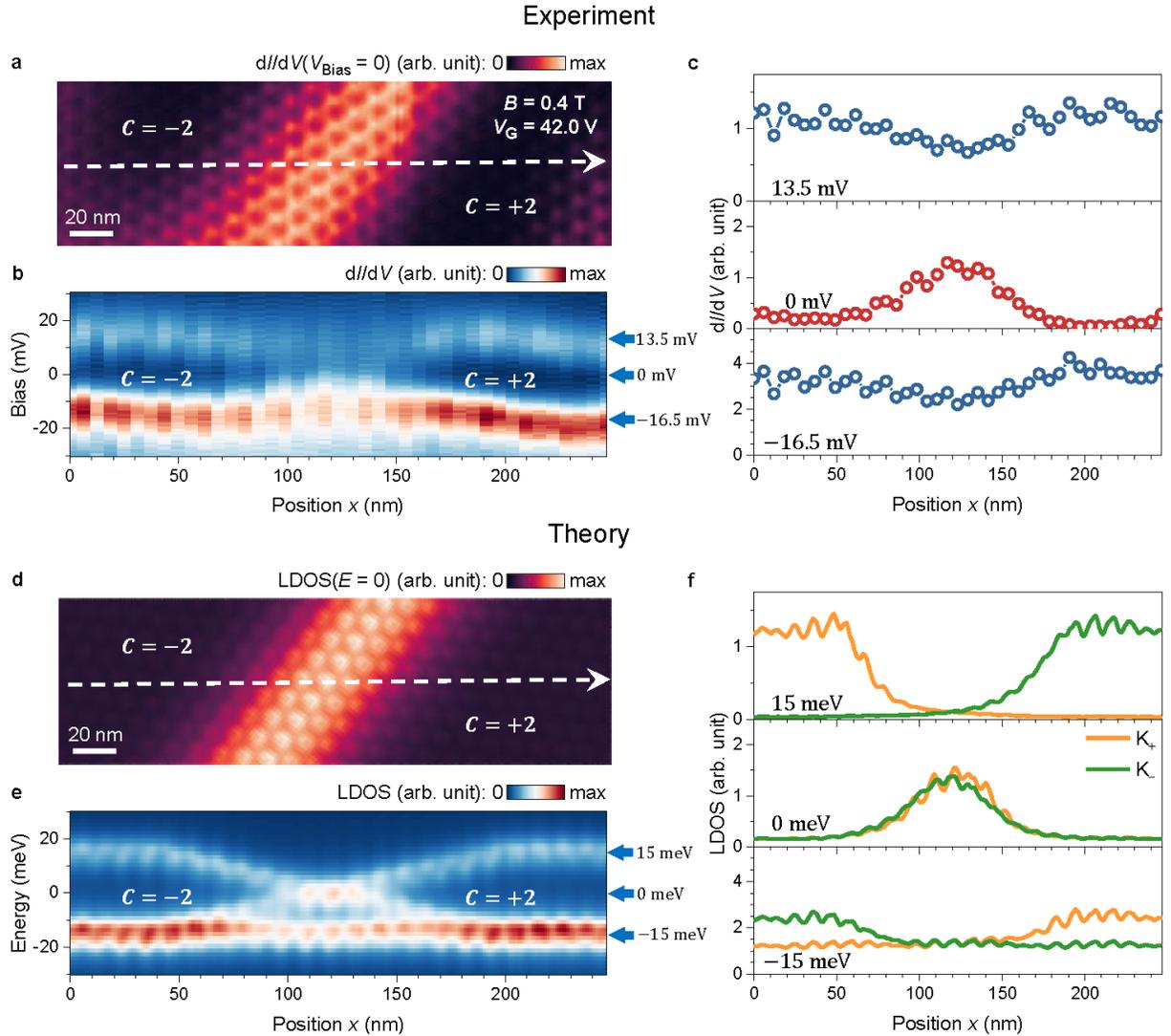

**Extended Data Figure 3: Experiment–theory comparison for topological phase transition and chiral interface states. a**, d$I$/d$V$ map of a tMBLG area for $B = 0.4$ T, $V_G = 42.0$ V, and $V_{Bias} = 0$ mV (same as Fig. 2e). **b**, d$I$/d$V$ density plot for $B = 0.4$ T and $V_G = 42.0$ V obtained along the white dashed line in **a** (same as Fig. 2f). **c**, d$I$/d$V$ spatial line-cuts extracted from **b** at $V_{Bias} = 13.5$ mV (top), 0 mV (middle), and –16.5 mV (bottom) (corresponding to blue arrows in **b**). **d**, Theoretical LDOS map for $E = 0$ meV calculated for a tMBLG Chern domain wall width of $\xi = 125$ nm. **e**, Density plot of theoretical LDOS along the white dashed line in **d** (same as Fig. 2g). **f**, Valley-resolved theoretical LDOS line-cuts at $E = 15$ meV (top), 0 meV (middle), and –15 meV (bottom) (corresponding to blue arrows in **e**).

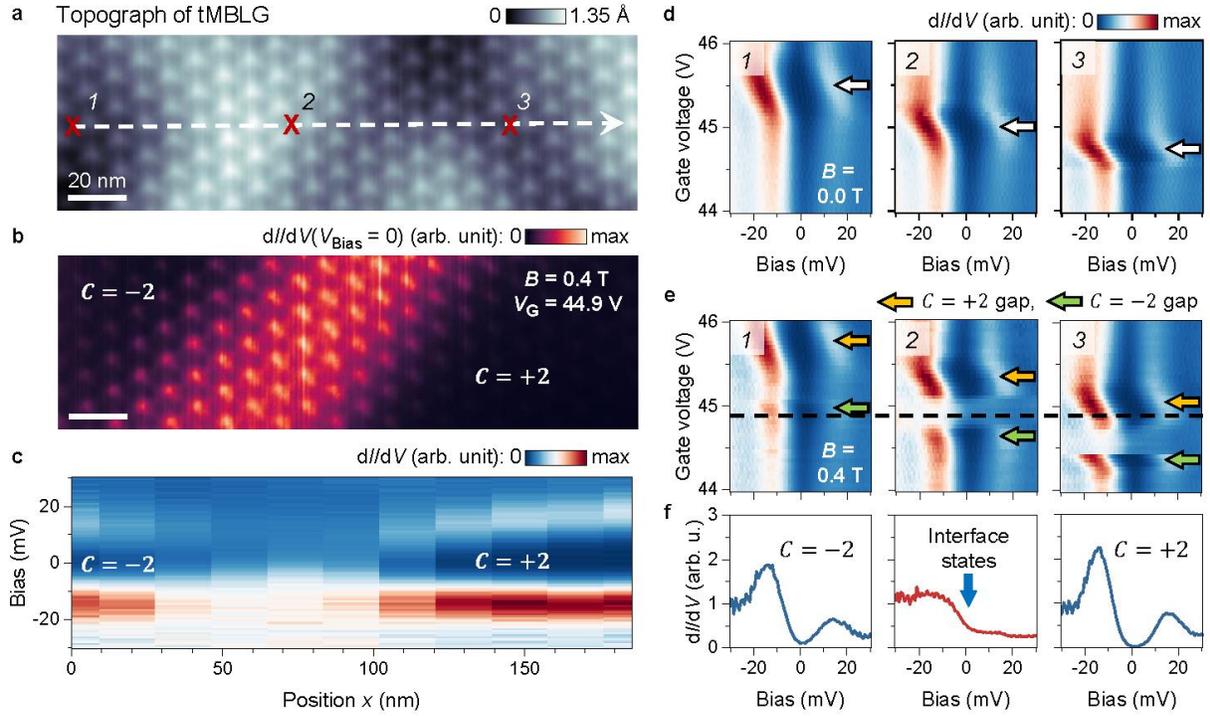

**Extended Data Figure 4: Spatially-defined topological phase transition and chiral interface states for a different area. a**, STM topographic image of a different tMBLG area ($V_{Bias} = -300$ mV, setpoint $I_0 = 0.2$ nA). **b**, d$I$/d$V$ map of the same area as **a** for $B = 0.4$ T, $V_G = 44.9$ V and $V_{Bias} = 0$ mV. **c**, d$I$/d$V$ density plot for $B = 0.4$ T and $V_G = 44.9$ V obtained along the white dashed line in **a**. **d**, Gate-dependent d$I$/d$V$ density plots for $B = 0.0$ T obtained at locations *1*, *2*, and *3* marked in **a**. White arrows indicate the $\nu = 3$ energy gap. **e**, Same as **d**, but for $B = 0.4$ T. Orange and green arrows indicate $C = +2$ and $C = -2$ gaps. **f**, d$I$/d$V$ spectra for $B = 0.4$ T and $V_G = 44.9$ V at locations *1*, *2*, and *3* of **a**. Spectroscopy parameters: modulation voltage $\Delta V_{RMS} = 1$ mV; setpoint $V_{Bias} = -60$ mV, $I_0 = 0.5$ nA for **d**, **e**; setpoint $V_{Bias} = -300$ mV, $I_0 = 0.2$ nA and tip height offset $\Delta Z = -0.2$ nm for **b**, **c**, **f**.

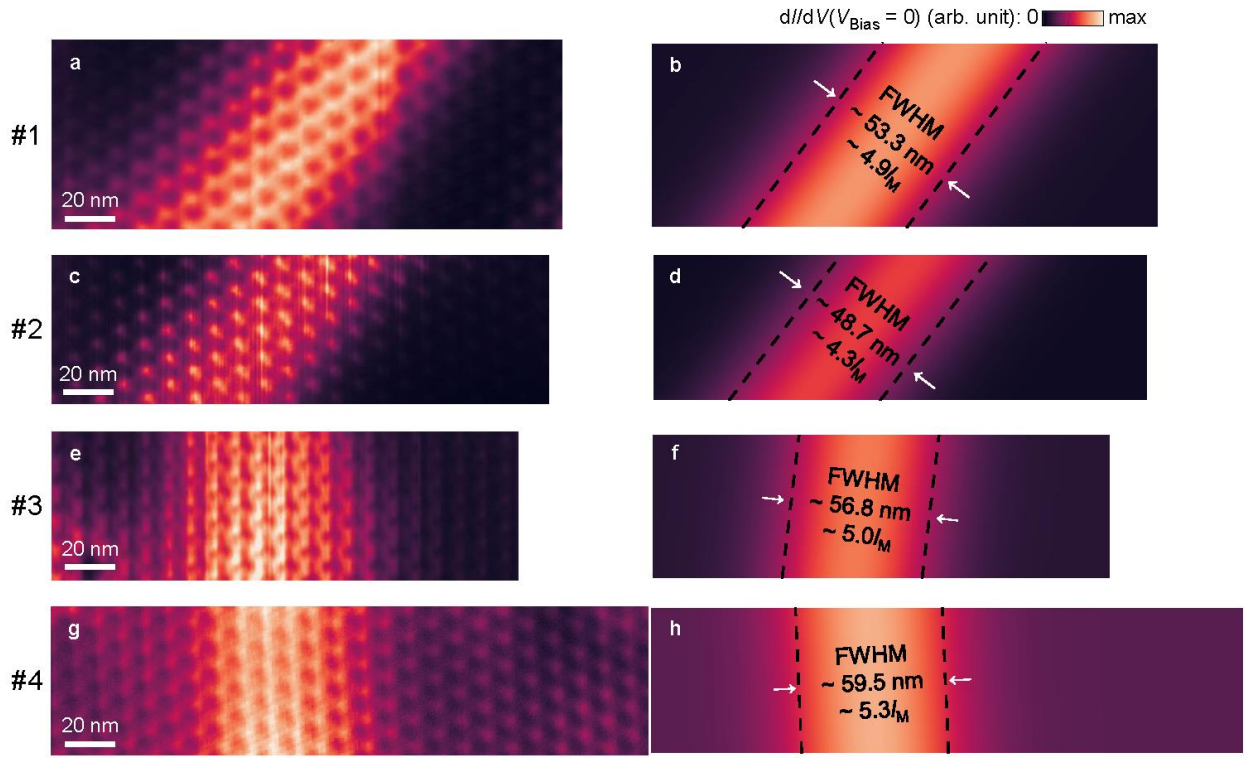

**Extended Data Figure 5: Fitting the spatial profile of chiral interface states. a**, Experimental d$I$/d$V$ map showing 1D chiral interface states obtained in area #1 for $B = 0.4$ T, $V_G = 42.0$ V and $V_{Bias} = 0$ mV (same as Fig. 2e). **b**, 2D fit to image in **a** with a Gaussian peak and a constant background (see Methods). **c**, Experimental d$I$/d$V$ map showing 1D chiral interface states obtained in area #2 for $B = 0.4$ T, $V_G = 44.9$ V and $V_{Bias} = 0$ mV (Same as Extended Data Fig. 4b). **d**, 2D fit to image in **c** with a Gaussian peak and a constant background (see Methods). **e**, Experimental d$I$/d$V$ map showing 1D chiral interface states obtained in area #3 for $B = 0.3$ T, $V_G = 42.0$ V and $V_{Bias} = 0$ mV. **f**, 2D fit to image in **e** with a Gaussian peak and a constant background (see Methods). **g**, Experimental d$I$/d$V$ map showing 1D chiral interface states obtained in area #4 for $B = 0.3$ T, $V_G = 42.1$ V and $V_{Bias} = 0$ mV. **h**, 2D fit to image in **g** with a Gaussian peak and a constant background (see Methods).

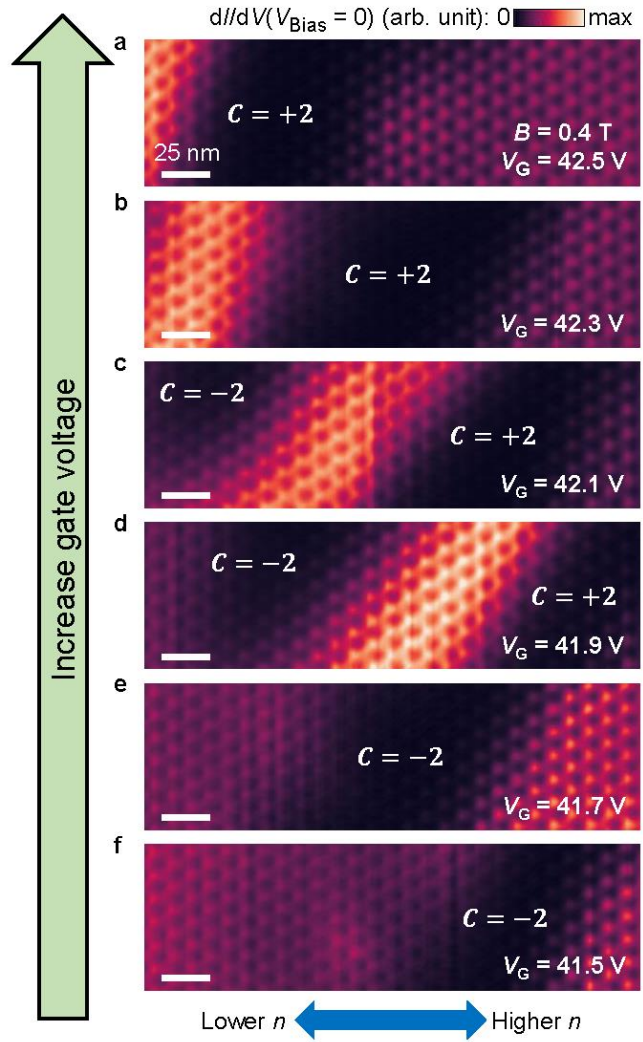

**Extended Data Figure 6: Reversible displacement of chiral interface states via back-gating.** **a–f**, d$I$/d$V$ maps of the same area as Fig. 2a at $B = 0.4$ T and $V_{Bias} = 0$ mV for **a** $V_G = 42.5$ V, **b** $V_G = 42.3$ V, **c** $V_G = 42.1$ V, **d** $V_G = 41.9$ V, **e** $V_G = 41.7$ V, and **f** $V_G = 41.5$ V. Modulation voltage $\Delta V_{RMS} = 1$ mV; setpoint $V_{Bias} = -300$ mV, $I_0 = 0.2$ nA; tip height offset $\Delta Z = -0.075$ nm.

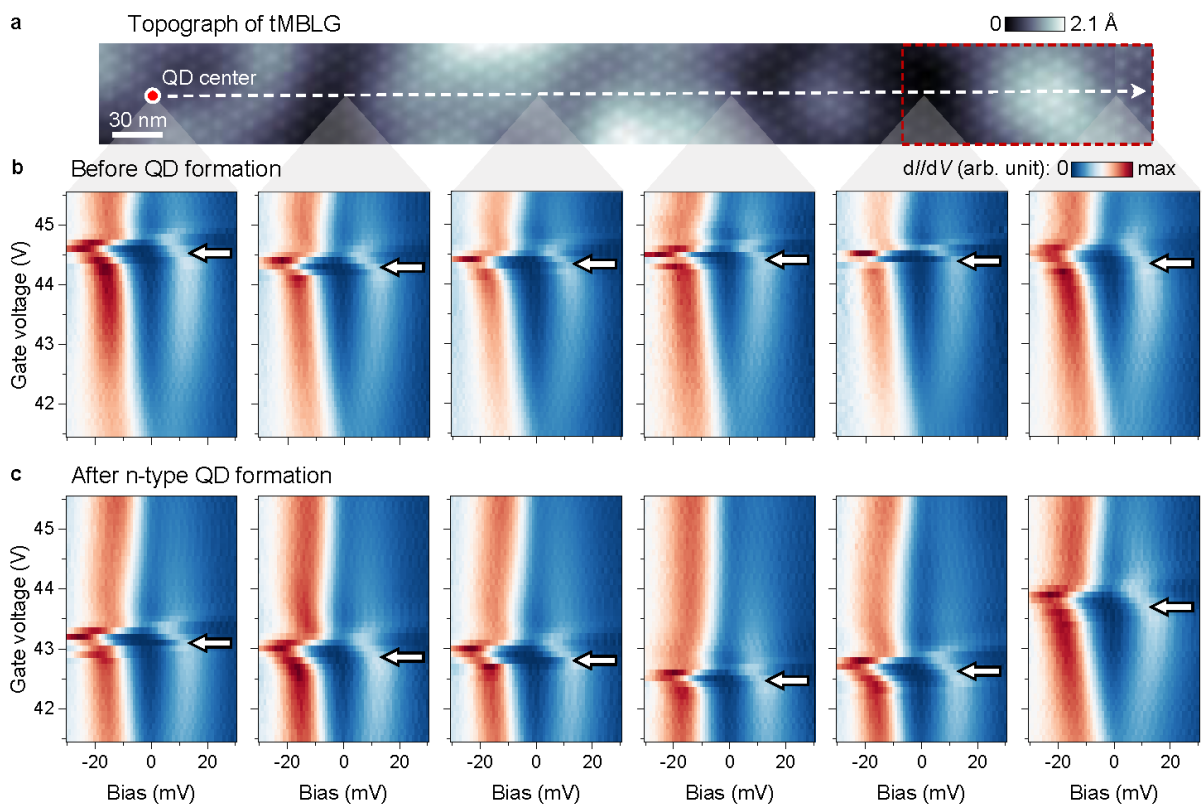

**Extended Data Figure 7: Local electron density profile of an n-type QD. a**, STM topographic image of a tMBLG area ($V_{Bias} = -1$ V, setpoint $I_0 = 0.01$ nA; same as Fig. 4a). **b**, Representative gate-dependent d$I$/d$V$ density plots for $B = 0.0$ T obtained at different locations indicated on white dashed line of **a**. The $\nu = 3$ gap (white arrows) always appears near $V_G = 44.3$ V, indicating a low variation of local electron density over the entire area in **a** ($\Delta\nu < 0.02$). **c**, Representative gate-dependent d$I$/d$V$ density plots for $B = 0.0$ T obtained at the same locations as in **b** after the formation of an n-type QD. A large electron density gradient occurs in the rightmost region (highlighted by the red dashed box in **a**) where a Chern domain interface hosting 1D chiral states can be realized (Fig. 4c). Spectroscopy parameters: modulation voltage $\Delta V_{RMS} = 1$ mV; setpoint $V_{Bias} = -60$ mV, $I_0 = 0.5$ nA.

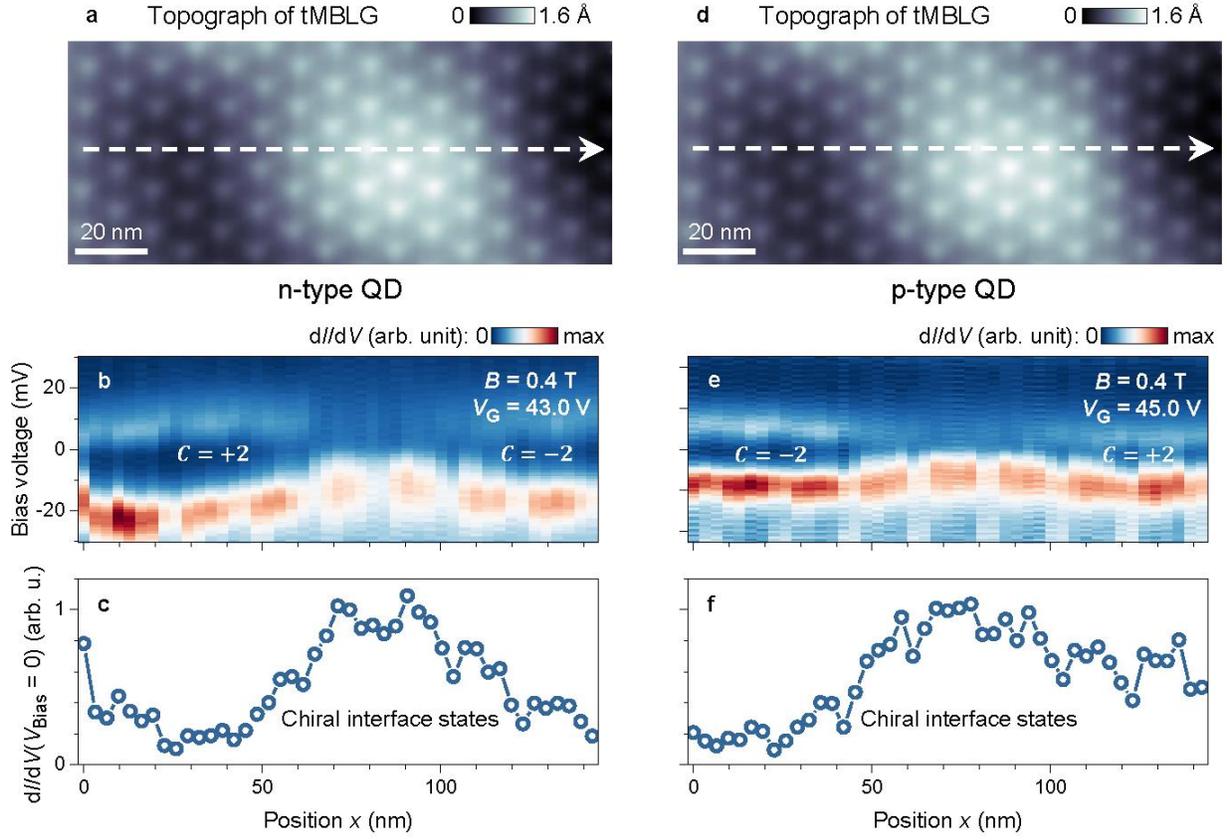

**Extended Data Figure 8: Spatially-defined topological phase transition and 1D chiral states at QD-created Chern domain interfaces. a**, Zoom-in STM topographic image of the boxed region shown in Fig. 4a ($V_{Bias} = -1$ V, setpoint $I_0 = 0.01$ nA). **b**, $dI/dV$ density plot for $B = 0.4$ T and $V_G = 43.0$ V obtained along the white dashed line in **a** after formation of an n-type QD (modulation voltage $\Delta V_{RMS} = 1$ mV; setpoint $V_{Bias} = -300$ mV, $I_0 = 0.2$ nA; tip height offset $\Delta Z = -0.075$ nm). Charge gaps in the left and right regions are attributed to the $C = +2$ and $C = -2$ insulating states. The absence of a gap in the middle signifies the presence of chiral interface states. **c**, $dI/dV$ spatial line-cut at $V_{Bias} = 0$ mV extracted from **b**. **d–f**, Same as **a–c**, but after formation of a p-type QD. Charge gaps in the left and right regions are attributed to the $C = -2$ and $C = +2$ insulating states. The absence of a gap in the middle signifies the presence of chiral interface states.

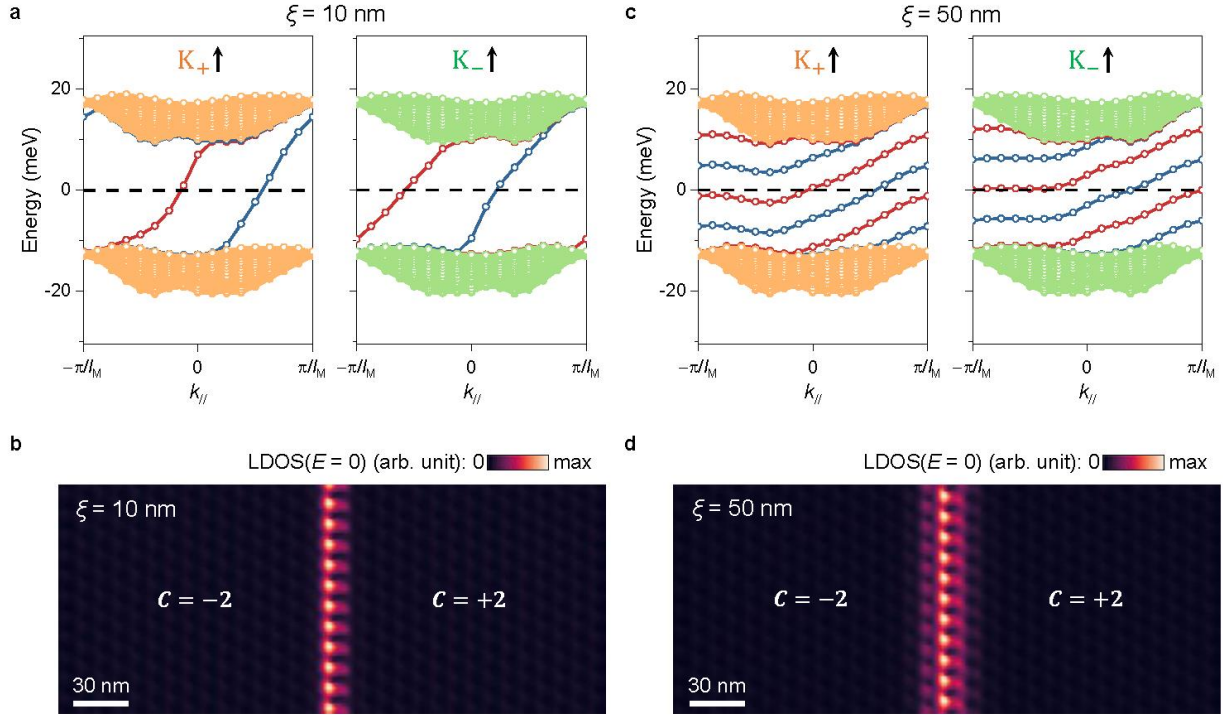

**Extended Data Figure 9: 1D band structure and chiral state image for a tMBLG Chern domain interface. a**, Energy eigenvalues as a function of momentum along the interface ($k_{\parallel}$) calculated for a Chern domain wall width of $\xi = 10$ nm (the minimum value allowed in our model). Only electronic states for spin up are shown (corresponding to the sub-bands in Fig. 2h that shift upward/downward). Two branches of chiral interface states (marked by red and blue) emerge for each valley that connect the occupied and unoccupied bulk bands (thus spanning the bulk gap) and disperse unidirectionally in momentum. The total number of chiral modes (four) is equal to the Chern number difference between neighboring domains. **b**, Theoretical LDOS map for $E = 0$ meV showing a FWHM of $w = 9$ nm for the 1D chiral interface modes arising from a Chern domain wall width of $\xi = 10$ nm. **c**, Energy eigenvalues calculated for a larger $\xi = 50$ nm. Here more in-gap eigenstates appear at a given momentum, but the number of chiral branches crossing $E_F$ (dashed lines) remains the same (two per valley, marked by red and blue). This is because the group velocity of interface modes becomes smaller (see Supplementary Note 3 for details) so they now extend over more than one Brillouin zone in momentum. **d**, Theoretical LDOS map for $E = 0$ meV showing a FWHM of $w = 17$ nm for the 1D chiral interface modes arising from a Chern domain wall width of $\xi = 50$ nm.

| Area | $n$-gradient ($10^8$ cm$^{-2}$/nm) | Origin of $n$-gradient | $B$ (T) | $V_G$ (V) | FWHM (nm) |
|---|---|---|---|---|---|
| #1 | 3.2 | Naturally occurring | 0.4 | 41.9 (inc.) | 51.7 |
| | | | | 42.0 (inc.) | 52.5 |
| | | | | 42.1 (inc.) | 56.2 |
| | | | | 42.1 (dec.) | 55.1 |
| | | | | 42.0 (dec.) | 53.3 |
| | | | | 41.9 (dec.) | 52.2 |
| #2 | 4.4 | Naturally occurring | 0.4 | 45.0 (dec.) | 51.5 |
| | | | | 44.9 (dec.) | 48.7 |
| | | | | 44.8 (dec.) | 46.9 |
| | | | | 44.7 (dec.) | 40.7 |
| #3 | 3.2 | STM tip pulse-induced QD | 0.4 | 42.0 (dec.) | 59.5 |
| | | | 0.3 | | 56.8 |
| #4 | 3.9 | STM tip pulse-induced QD | 0.3 | 42.1 (dec.) | 59.5 |
| | | | 0.2 | | 66.6 |

**Extended Data Table 1: Summary of 2D Gaussian fitting for different datasets shown in Extended Data Fig. 5.** inc. = increasing gate voltage; dec. = decreasing gate voltage. Areas #1–4 can be seen in Extended Data Fig. 5.